\documentclass[prd,twocolumn,nofootinbib,showpacs,eqsecnum,floatfix,superscriptaddress]{revtex4}
\usepackage{latexsym,amsbsy}
  \usepackage[dvips]{graphicx}
 \DeclareGraphicsExtensions{.eps, .jpg}

\begin{document}
\title{String Models, Stability  and Regge Trajectories for Hadron States}
%
\author{German Sharov}
\email{german.sharov@mail.ru} \affiliation{Tver state university, 170002, Sadovyj per.
35, Tver, Russia}
%

\def\s{\sigma}
\def\al{\alpha}
\def\lm{\lambda}
\def\de{\delta}
\def\om{\omega}
\def\pr{\prime}
\def\f{\varphi}

\begin{abstract}
Various string  models of mesons and baryons include a string
carrying 2 or 3 massive points (quarks or antiquarks). Rotational
states (planar uniform rotations) of these systems generate
quasilinear Regge trajectories and may be used for describing
excited hadron states on these trajectories. For different string
models of baryon we are to solve the problem of choice between
them and the stability problem for their rotational states. An
unexpected result is that for the Y string baryon model these
rotations are unstable with respect to small disturbances on the
classical level. This instability has specific feature,
disturbances grow linearly, whereas for the linear string baryon
model they grow exponentially and may increase predictions for
baryon's width $\Gamma$.

The classical instability of rotational states and nonstandard
Regge slope are the arguments in favor of the stable simplest
model of string with massive ends both for baryons and mesons.
Rotational states of this model with two types of spin-orbit
correction are used to describe Regge trajectories for light,
strange, charmed, bottom mesons and for $N$, $\Delta$, $\Sigma$,
$\Lambda$ and $\Lambda_c$  baryons.

\vspace{5mm}


\end{abstract}

\maketitle

\section{Introduction}\label{Intr}
\begin{figure}[t]
\includegraphics[scale=0.8,trim=6mm 5mm 5mm 2mm]{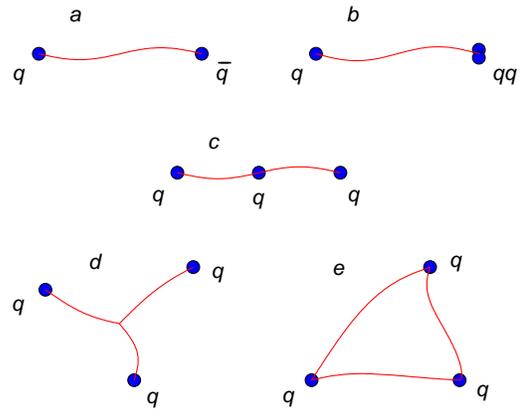}
\caption{String models of mesons and baryons}
\label{mod}\end{figure}

 In string models of mesons and baryons
\cite{Nambu,Ch,AY,4B,Ko,PRTr,InSh,Solovm,stab,PR09,Y10} shown in
Fig.~\ref{mod} the Nambu-Goto string (relativistic string)
simulates strong interaction between quarks at large distances and
QCD confinement mechanism. This string has linearly growing energy
with constant energy density equal to the string tension $\gamma$.

Such a string with massive ends \cite{Ch} may be regarded as the
meson string model in Fig.~\ref{mod}{\it a} or the quark-diquark
model $q$-$qq$ \cite{Ko} in Fig.~\ref{mod}{\it b}  (on the classic
level these models coincide). Other string models of baryons are
\cite{AY}:  ({\it c}) the linear configuration $q$-$q$-$q$
\cite{PR09}, ({\it d}) the ``three-string'' model or Y
configuration \cite{AY,stab,Y10}, and (e) the ``triangle'' model
or $\Delta$ configuration \cite{PRTr,stab}.

For all cited string hadron models one can use rotational states
of these systems (classical planar uniform rotations) to describe
 quasilinear Regge trajectories for mesons and baryons
\cite{4B,Ko,PRTr,InSh,Solovm}. In the limit of large energies $E$
for a rotational state the angular momentum $J$ of this state
behaves as
 $ J\simeq\al'E^2$ for any model in Fig.~\ref{mod}.
 For the meson and baryon models in
Fig.~\ref{mod}{\it a}, {\it b} and {\it c} the slope $\al'$ and
the string tension $\gamma$ are connected by  Nambu relation
\cite{Nambu} $ \alpha'=(2\pi\gamma)^{-1}$. So, if we use these
models with the same type of strings (the fundamental string), we
can naturally describe baryonic and mesonic Regge trajectories
with the same experimental slope $\al'\simeq0{.}9$ GeV$^{-2}$.

Rotational states of the string baryon model Y
(Fig.\,\ref{mod}{\it d}) demonstrate the Regge asymptotics with
the slope \cite{4B} $\alpha'=1/(3\pi\gamma)$. To obtain the
experimental value $\al'\simeq0{.}9$ GeV$^{-2}$  we are to assume
that the effective string tension $\gamma_Y$ in this model differs
from the fundamental string tension $\gamma$ in models in
Figs.\,\ref{mod}{\it a\,--\,c}  and equals
$\gamma_Y=\frac23\gamma$ \cite{4B,InSh}.

The string baryon model ``triangle'' or $\Delta$ encounters the
similar problem. For describing Regge trajectories with the so
called triangle rotational states \cite{PRTr}  we are to take
another effective string tension $\gamma_\Delta=\frac38\gamma$
\cite{4B,InSh}.

 To choose the most adequate string model
of a baryon one should analyze the stability problem for
rotational states of these models. Stability of classical
rotational states with respect to small disturbances for the
models in Fig.\,\ref{mod} was studied in numerical experiments
\cite{stab} and analytically \cite{PR09,Y10}. Rotational states
for some models, in particular, for the linear model and the Y
configuration appeared to be unstable. This fact is very important
for applications of these models in hadron spectroscopy.

 Note that instability of classical rotations for
some string configuration does not mean that the considered string
model must be totally prohibited. All excited hadron states
(objects of modelling) are resonances, they are unstable with
respect to strong decays. So they have rather large width
$\Gamma$. If classical rotations of a string configuration are
unstable and this instability has a characteristic time scale
$t_{inst}$, it gives the additional contribution
$\Gamma_{inst}\simeq1/t_{inst}$ to width $\Gamma$. This effect can
restrict applicability of some string models, if the value
$\Gamma_{inst}$ predicted by this model essentially exceeds
experimental data for $\Gamma$ \cite{PR09}.

In this paper we describe dynamics of the mentioned string hadron
models and the stability problem for their rotations in
Sect.~\ref{Dyn}. In Sect.~\ref{Regge} string models with stable
rotational states are applied to Regge trajectories for mesons and
baryons.

\section{Dynamics and stability of string hadron models}\label{Dyn}

Dynamics of an open or closed string carrying $n=2$  or 3
point-like masses $m_1$, $m_2,\;\dots\,m_n$ (the models in
Figs.\,\ref{mod}{\it a\,--\,c} or {\it e}) is determined by the
action \cite{4B,PRTr,PR09}
 \begin{equation}
A=-\gamma\int\limits_{D}\sqrt{-g}\;d\tau d\s -\sum\limits_{j=1}^n
m_j\int\sqrt{\dot x_j^2(\tau)}\;d\tau.
 \label{S}\end{equation}
 Here $\gamma$ is the string tension, $g$ is the determinant of the
induced metric
$g_{ab}=\eta_{\mu\nu}\partial_aX^\mu\partial_bX^\nu$ on the string
world surface $X^\mu(\tau,\s)$ embedded in Minkowski space
$R^{1,3}$, $\eta_{\mu\nu}=\mbox{diag}(1,-1.-1,-1)$, the speed of
light $c=1$, the domain
 $
 D=\big\{(\tau,\s):\,\tau\in R,\;\s_1(\tau)<\s<\s_n(\tau)\big\}
 $ for the models in
Figs.\,\ref{mod}{\it a\,--\,c} (or with $\s_0<\s<\s_n$ for the
closed string), world lines of massive points are
$$x_j^\mu(\tau)=X^\mu(\tau,\s_j(\tau)),\quad j=1,\dots,n.$$

Equations of motion for these  string models result from the
action (\ref{S}). Without loss of generality \cite{4B,PRTr}  we
choose coordinates $\tau$, $\s$ satisfying the orthonormality
conditions on the world surface
 \begin{equation}
(\partial_\tau X\pm\partial_\s X)^2=0.
 \label{ort}\end{equation}
 Under these conditions  the equations of motion are reduced to the
 string motion equation
 \begin{equation} \frac{\partial^2X^\mu}{\partial\tau^2}-
\frac{\partial^2X^\mu}{\partial\s^2}=0,
 \label{eq}\end{equation}
 and equations for two types of massive points: for endpoints
 \begin{equation}
 m_j\frac d{d\tau}\frac{\dot
x_j^\mu(\tau)}{\sqrt{\dot x_j^2(\tau)}}+\epsilon_j
\gamma\bigl[X^{'\!\mu}+\dot\s_j(\tau)\,\dot
X^\mu\bigr]\bigg|_{\s=\s_j}=0 \label{qq1}
 \end{equation}
 (here $\epsilon_1=-1$, $\epsilon_n=1$) and for the middle point in the model $q$-$q$-$q$ or for all points on a
 closed string
 \begin{eqnarray} m_j\frac d{d\tau}\frac{\dot
x_j^\mu(\tau)}{\sqrt{\dot x_j^2(\tau)}}+\gamma
\Big[X^{'\!\mu}+\dot\s_j(\tau)\dot
X^\mu\Big]\Big|_{\s=\s_j-0}&&\nonumber\\
{}-\gamma\Big[X^{'\!\mu}+\dot\s_j(\tau)\dot
X^\mu\Big]\Big|_{\s=\s_j+0}=0,&&
 \label{qqi}\end{eqnarray}
 Here $\dot X^\mu\equiv\partial_\tau X^\mu$,
$X^{'\!\mu}\equiv\partial_\s X^\mu$, the scalar product
$(\xi,\zeta)=\eta_{\mu\nu}\xi^\mu\zeta^\nu$.

For the open string with $n=2$ or 3 masses (the models in
Figs.\,\ref{mod}{\it a\,--\,c}) rotational states are uniform
rotations of a rectilinear string segment. The correspondent
solution of Eqs.~(\ref{ort})\,--\,(\ref{qqi}) may be presented in
the form  \cite{stab,PR09}:
 \begin{equation}
 X^\mu(\tau,\s)
 =\Omega^{-1}
\big[\theta\tau e_0^\mu+\cos(\theta\s+\phi_1)\cdot
e^\mu(\tau)\big].
 \label{lrot}\end{equation}
 Here we fixed conditions at the ends in
Eqs.~(\ref{qq1}) \cite{4B}:
 \begin{equation}
 \s_1(\tau)=0,\qquad \s_n(\tau)=\pi,\qquad \s\in[0,\pi],
\label{s0pi}\end{equation}
 $\Omega$ is the angular velocity, $e_0,\,e_1,\,e_2,\,e_3$ is
the orthonormal tetrade in Minkowski space $R^{1,3}$,
 \begin{equation}
e^\mu(\tau)=e_1^\mu\cos\theta\tau+e_2^\mu\sin\theta\tau
\label{e}\end{equation}
 is the
unit rotating vector directed along the string. Values $\theta$
(dimensionless frequency) and $\phi_1$ are connected with the
constant speeds $v_j$ of the ends
 $$ 
  v_1=\cos\phi_1,\;\;\;  v_n=-\cos(\pi\theta +\phi_1),\;\;\;
\frac{m_j\Omega}\gamma=\frac{1-v_j^2}{v_j}.
 $$ 
In the $q$-$q$-$q$ system the central massive point is at rest (in
the comoving frame) at the rotational center. Its inner coordinate
is
 $ \s_2=(\pi-2\phi_1)/(2\theta)={\mbox{const}}.$

In Refs.~\cite{stab,qrottmf} we analyzed stability of the
rotational states (\ref{lrot}) for the string with massive ends
(Fig.~\ref{mod}{\it a, b}). These states appeared to be stable
with respect to small disturbances.

For the linear string baryon model $q$-$q$-$q$  the stability
problem for the states (\ref{lrot}) was solved in numerical
experiments \cite{stab} and analytically \cite{PR09}. Analysis in
Ref.~\cite{PR09} demonstrated that the rotational states
(\ref{lrot}) for this system are unstable, because an arbitrary
disturbed rotation has complex (imaginary) frequencies in its
spectrum. They correspond to exponentially growing modes of  small
disturbances:
 \begin{equation}
 \delta X^\mu\sim\exp(t/t_{inst}).
 \label{expo}\end{equation}
 Calculations of the characteristic time $t_{inst}$ and its reciprocal
$\Gamma_{inst}\simeq1/t_{inst}$ in Ref.~\cite{PR09} showed that
the value $\Gamma_{inst}$ for the model $q$-$q$-$q$ strongly
depends on energy $E$ of the rotational state. We are to compare
this value with the experimental width $\Gamma$ corresponding to
strong decays of a baryon. In string models these decays are
described as string breaking with probability or width
$\Gamma=\Gamma_{br}$, proportional to the string length $\ell$
\cite{Kodecay,GuptaR}.

In Ref.~\cite{PR09} we estimated the total width, predicted by the
baryon model $q$-$q$-$q$ as $\Gamma=\Gamma_{br}+\Gamma_{inst}$ in
comparison with experimental data for $N$, $\Delta$ and strange
baryons in the mass (or energy) range 1 -- 3 GeV. For $E\simeq1$
 GeV the contribution $\Gamma_{inst}$ in total width $\Gamma$
 appeared to be essentially exceeding experimental data.
 So we concluded, that the linear string model
$q$-$q$-$q$ is unacceptable for describing these baryon states and
we should refuse this model in favor of the quark-diquark or Y
models.

For the string baryon model Y (Fig.~\ref{mod}{\it d}) three world
sheets (swept up by three string segments) are parametrized with
three different functions $X_j^\mu(\tau_j,\s)$ \cite{stab,Y10}. It
is convenient to use different notations $\tau_1$, $\tau_2$,
$\tau_3$ for ``time-like'' parameters and the same symbol $\s$ for
``space-like'' parameters. These three world sheets are joined
along the world line of the junction that may be set as $\s=0$ for
all sheets without loss of generality. At this junction parameters
$\tau_j$ are connected as follows~\cite{stab}
$$\tau_2=\tau_2(\tau),\quad\tau_3=\tau_3(\tau),\quad\tau_1\equiv\tau.$$
 So at the junction we have the condition
 \begin{equation}
X_1^\mu\big(\tau,0\big)=X_2^\mu\big(\tau_2(\tau),0\big)=
X_3^\mu\big(\tau_3(\tau),0\big). \label{junc}\end{equation}

The action of the Y configuration \cite{stab,Y10} looks like
Eq.~(\ref{S}), but the first term includes three integrals along
the mentioned world sheets. So dynamical equations for this model
include the same equations (\ref{eq})
 \begin{equation}
\frac{\partial^2X_j^\mu}{\partial\tau_j^2}-
\frac{\partial^2X_j^\mu}{\partial\s^2}=0
 \label{eqy}
 \end{equation}
 under the  conditions (\ref{ort})
$
 (\partial_{\tau_j} X_j\pm\partial_\s X_j)^2=0
$ on three world sheets and also conditions (\ref{s0pi})
$0\le\s\le\pi$ and equations  (\ref{qq1}) for massive endpoints
with $\epsilon_j=1$,  $\s_j=\pi$ for all $j=1,2,3$.
 One should substitute $\tau\to\tau_j$, $X^\mu\to X_j^\mu$ in Eq.~(\ref{qq1}) and
add the relation at the junction
 \begin{equation}
\sum_{j=1}^3 X_j^{'\!\mu}\big(\tau_j(\tau),0\big)\, \frac
{d\tau_j(\tau)}{d\tau}=0.  
\label{qy}
 \end{equation}

Rotational states of the  Y configuration correspond to planar
uniform rotation of three rectangular string segments connected at
the junction at angles of $120^\circ$ \cite{4B,Ko,stab}. These
states may be described as Eq.~(\ref{lrot}) \cite{Y10}
 \begin{equation}
 \underline X_j^\mu(\tau_j,\s)= \Omega^{-1}\big[\theta\tau_j
e_0^\mu+\sin(\theta\s)\cdot e^\mu(\tau_j+\Delta_j)\big].
 \label{roty}\end{equation}
  Here $\tau_1=\tau_2=\tau_3$,
$\Delta_j=2\pi(j-1)/(3\theta)$, $ e^\mu(\tau)$
 is the unit rotating vector (\ref{e}) directed along the first
string segment. Below we consider the symmetric case \cite{Y10}
 \begin{equation}
 m_1=m_2=m_3,\qquad v_1=v_2=v_3
 \label{m123}\end{equation}

Expression (\ref{roty}) satisfies Eq.~(\ref{eqy}) and conditions
(\ref{ort}), (\ref{qq1}), (\ref{junc}), (\ref{qy}), if angular
velocity $\Omega$, the value $\theta$, constant velocities $v_j$
of the massive points are connected by the relations \cite{4B}
 \begin{equation}
v_j=\sin(\pi\theta)=\left[\Big(\frac{\Omega
m_j}{2\gamma}\Big)^2+1\right]^{1/2}- \frac{\Omega m_j}{2\gamma}.
\label{v1}\end{equation}

In Ref.~\cite{stab} we demonstrated in numerical experiments, that
rotational states (\ref{roty}) of the Y configuration are unstable
with respect to small disturbances. This instability was
investigated by G. 't Hooft \cite{tHooft04}. Here we test this
stability problem analytically.

Let us consider a slightly disturbed motion of the model Y with a
world surface $X_j^\mu(\tau_j,\s)$ close to the surface
$\underline X_j^\mu(\tau_j,\s)$ of the rotational state
(\ref{roty}) (below we underline values, describing rotational
states). For this disturbed motion we use the general solution
of Eq.~(\ref{eqy})
 \begin{equation}
 X_j^\mu(\tau_j,\s)=\frac1{2}\bigl[\Psi^\mu_{j+}(\tau_j+\s)+
\Psi^\mu_{j-}(\tau_j-\s)\bigr],
 \label{soly}\end{equation}
 for every world sheet. Functions
$\Psi^\mu_{j\pm}(\tau)$ have isotropic derivatives with respect to
their argumetns
 \begin{equation}
 \dot\Psi_{j+}^2=\dot\Psi_{j-}^2=0.
 \label{isotr}\end{equation}
 as a consequence of the orthonormality conditions (\ref{ort}).

If we substitute Eq.~(\ref{soly}) into conditions (\ref{junc}),
(\ref{qy}) and (\ref{qq1}), they may be reduced to the form
\cite{Y10}
\begin{eqnarray}
& \Psi^\mu_{1+}(\tau)+\Psi^\mu_{1-}(\tau)=
 \Psi^\mu_{j+}(\tau_j)+\Psi^\mu_{j-}(\tau_j), &\label{jps1}\\
& \sum\limits_{j=1}^3\big[\dot\Psi^\mu_{j+}(\tau_j)-
\dot\Psi^\mu_{j-}(\tau_j)\big]\dot\tau_j(\tau)=0,&
 \label{jps2}\\
&\displaystyle\dot\Psi_{j\pm}^\mu(\tau_j\pm\pi)=
 \frac{m_j}{\gamma}\Big[\sqrt{-\dot
U_j^2 (\tau_j)}\,U_j^\mu(\tau_j)\mp\dot
U_j^\mu(\tau_j)\Big].\qquad&
 \label{PsUy}
\end{eqnarray}
  Here
 $$ 
 U^\mu_j(\tau_j)= 
 \frac{\dot\Psi_{j+}^\mu(\tau_j+\pi)+\dot\Psi_{j-}^\mu(\tau_j-\pi)}
{\big[2\big(\dot\Psi_{j+}(\tau_j+\pi),\dot\Psi_{j-}(\tau_j-\pi)\big)\big]^{1/2}}
 $$ 
 are velocities of massive ends.
Equations (\ref{PsUy}) and (\ref{soly}) determine functions
$X_j^\mu(\tau_j,\s)$ for world sheets if we know velocities
$U_j^\mu(\tau_j)$. So we search vectors
 $U_j^\mu$ for disturbed motion as small corrections to
velocities $\underline U_j^\mu$ for rotational states
(\ref{roty}):
 \begin{equation}
U_j^\mu(\tau_j)= \underline U_j^\mu(\tau_j)+u_j^\mu(\tau_j).
 \label{U+u}\end{equation}
 Here the vectors $\underline U_j^\mu$ are
 \begin{equation}
 \underline U_j^\mu(\tau_j)= (1-v_j^2)^{-1/2} 
 \big[e_0^\mu+v_j\acute e^\mu(\tau_j+\Delta_j)
\big],\label{Uroty}\end{equation}
 rotating vector
 $\acute e^\mu(\tau)=-e_1^\mu\sin(\theta\tau)+e_2^\mu\cos(\theta\tau)$
 is orthogonal to $e^\mu(\tau)$ (\ref{e}).

We suppose that for a disturbed motions the ``time'' parameters
 \begin{equation}
\tau_j(\tau)=\tau+\de_j(\tau),\quad j=2,3,\quad|\de_j(\tau)|\ll1,
 \label{taude}\end{equation}
 have small deviations $\de_2(\tau)$ and $\de_3(\tau)$ from $\tau\equiv\tau_1$,
 and also suppose disturbances $u_j^\mu(\tau_j)$ in Eq.~(\ref{U+u}) to be small
 ($|u_j^\mu|\ll1$).
So we omit squares of $u_j$ and $\de_j$ when we substitute the
expressions (\ref{U+u}) and (\ref{taude}) for disturbed motion
into dynamical equations (\ref{jps1})\,--\,(\ref{PsUy}) and into
equalities $U_j^2=\underline U_j^2=1$, resulting in relations
 \begin{equation}
\big(\underline U_j(\tau_j),u_j(\tau_j)\big)=0.
 \label{Uu}\end{equation}

After this substitution we have the linearized system (with
respect to $u_j^\mu$, $\de_j$) including Eqs.~(\ref{Uu}) and the
following vector equations:
 \begin{equation}
\begin{array}{c}
 \makebox[6cm]{$\sum\limits_\pm\Big[Qu_1^\mu(\pm)\pm\dot
u_1^\mu(\pm)+\underline
U_1^\mu(\pm)\big(e(\pm),\dot u_1(\pm)\big)\Big]=$}\\
\makebox[6cm]{$=\sum\limits_\pm\Big\{Qu_j^\mu(\pm)\pm\dot
u_j^\mu(\pm)+\underline U_j^\mu(\pm)\big(e(\pm_j),\dot
u_j(\pm)\big)+$}\\ \displaystyle
+\frac\gamma{m_1}\Big[\dot\de_j(\tau)\dot{\underline\Psi}_{j\pm}^\mu(\tau)+
\de_j(\tau)\ddot{\underline\Psi}_{j\pm}^\mu(\tau)\Big]\Big\},\\
 \displaystyle
 \sum\limits_{j=1}^3\sum\limits_\pm\Big\{\mp
 \frac\gamma{m_1}\Big[\dot\de_j\dot{\underline\Psi}_{j\pm}^\mu(\tau)+
\de_j\ddot{\underline\Psi}_{j\pm}^\mu(\tau)\Big]+\\
 \makebox[6cm]{$+ \dot u_j^\mu(\pm)\pm Qu_j^\mu(\pm)\pm\underline
U_j(\pm)\big(e(\pm_j),\dot u_j(\pm)\big)\Big\}=0.$}
 \end{array}
 \label{sysu}\end{equation}
 Here $(\pm)\equiv(\tau\pm\pi)$,
 $(\pm_j)\equiv(\tau\pm\pi+\Delta_j)$, the functions
 \begin{equation}
 \dot{\underline\Psi}_{j\pm}^\mu(\tau)=\frac{m_1Q}{\gamma
 c_1} \big[e_0^\mu+v_1\acute e^\mu(\mp_j)\pm c_1e^\mu(\mp_j)
 \big]
 \label{Psrot}\end{equation}
 correspond to rotational state (\ref{roty}),
 \begin{equation}
Q=\theta v_1/c_1,\qquad c_1=\cos(\pi\theta)=\sqrt{1-v_1^2}.
 \label{Q}\end{equation}

We search oscillatory solutions of this system  and substitute the
following disturbances with $u_j^\mu$ satisfying Eqs.~(\ref{Uu})
 \begin{eqnarray}
u_j^\mu(\tau)&=&\big[A_j^0e_0^\mu+A_j^ze_3^\mu+c_1A_je^\mu(\tau+\Delta_j)+{}\nonumber\\
&+&v_1^{-1}A_j^0 \acute
e^\mu(\tau+\Delta_j)\big]\exp(-i\xi\tau), \label{uA}\\
 \de_j(\tau)&=&\de_j\exp(-i\xi\tau),\quad j=2,3
 \label{dej}
 \end{eqnarray}
 into the system (\ref{sysu}).
Projections of these equations onto basis vectors $e_0^\mu$,
$e^\mu(\tau)$, $\acute e^\mu(\tau)$, $e_3^\mu$ form the system of
algebraic equations with respect to the small complex amplitudes
$A_j^0$, $A_j$, $A_j^z$, $\de_j$.

These projections onto the vector $e_3^\mu$ are
 $$\begin{array}{c}
 (\xi\tilde c +Q\tilde s)(A_1^z+A_2^z+A_3^z)=0,\\
 (\xi\tilde s-Q\tilde c)(A_1^z-A_j^z)=0,\quad j=2,3.\rule{0mm}{1.4em}
 \end{array}$$
 Here $\,\tilde c=\cos\pi\xi$, $\,\tilde s=\sin\pi\xi.$
 Solutions of these equations
describe 2 types of small oscillations of rotating Y configuration
(in $e_3$-direction). Corresponding frequencies $\xi$ of these
oscillations are roots of the equations
 \begin{equation}
 \xi/Q=\cot\pi\xi,\qquad
\xi/Q=-\tan\pi\xi.
 \label{zfreqy}\end{equation}
 All roots of Eqs.~(\ref{zfreqy}) are simple roots and real numbers,
therefore amplitudes of such fluctuations do not grow.


Small disturbances in the rotational plane ($e_1,e_2$) are
described by projections of Eqs.~(\ref{sysu}) onto 3 vectors
$e_0$, $e(\tau)$, $\acute e(\tau)$. These projections form the
system of 9 linear equations (with 8 independent ones among them)
with respect to 8 unknown values $A_j^0$, $A_j$, $\de_j$.
Nontrivial solutions of this system exist if and only if its
determinant equals zero. This equality after simplification,  is
reduced to the following equation Ref.~\cite{Y10}:
 \begin{equation}
(\xi^2-\theta^2)\bigg(\frac{\xi^2-q}{2Q\xi}+\tan\pi\xi\bigg)\bigg(
\frac{\xi^2-q}{2Q\xi}-\cot\pi\xi\bigg).
 \label{pfr1}\end{equation}
 Here $q=Q^2(1+v_1^{-2})=\theta^2(1+v_1^2)/(1-v_1^2)$.


We analyzed roots of this equation  for complex values
$\xi=\xi_1+i\xi_2$ in Ref.~\cite{Y10} and concluded that all roots
of Eq.~(\ref{pfr1}) are real numbers and form a countable set.
This behavior differs from that for the linear string model
$q$-$q$-$q$. In the latter case the corresponding spectral
equation has complex roots (frequencies of disturbances)
\cite{PR09}, so these disturbances grow exponentially in
accordance with Eq.~(\ref{expo}).

For the model Y the observed in Ref.~\cite{stab} instability of
rotational states (\ref{roty}) has another nature. This
instability results from existence of double roots $\xi=\pm\theta$
in Eq.~(\ref{pfr1}).  If we put $\xi=\pm\theta$ in
Eq.~(\ref{pfr1}), not only the first factor $(\xi^2-\theta^2)$,
but also the second factor vanishes:
 $$
\frac{\theta^2-q}{2Q\theta}+\tan\pi\theta=0.
 $$
 This equality results from Eqs.~(\ref{v1}),
(\ref{Q}).

Double roots of Eq.~(\ref{pfr1}) correspond to oscillatory modes
with linearly growing amplitude. If we fix frequency $\xi=\theta$
and substitute small disturbances in the form (\ref{uA}),
(\ref{dej}) but with $A_j+\tilde A_j\tau$ and $\de_j+\tilde \de
A_j\tau$ instead of $A_j$ and $\de_j$ into Eqs.~(\ref{sysu}), we
can find nontrivial solutions with linearly growing amplitude:
 \begin{equation}
 |u_j^\mu|\simeq|\tilde A_1|c_1\tau\qquad
 |\de_j|\simeq \sqrt3c_1^2Q_1^{-1}|\tilde A_1|\tau. \label{linmod}
 \end{equation}

\begin{figure}[bh]
\includegraphics[scale=0.63,trim=15 5 10 20]{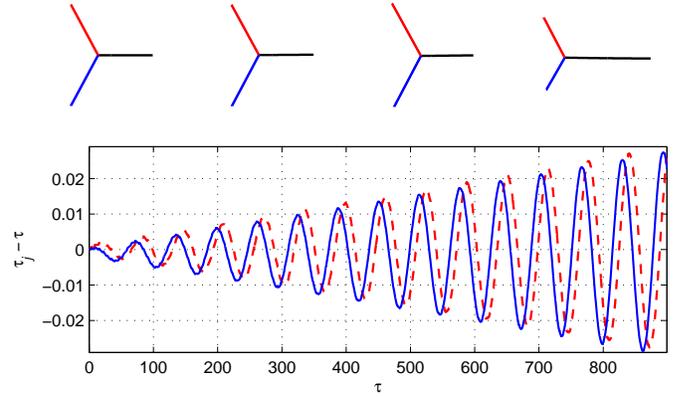}
\caption{Shape of the Y configuration and dependence of
$\tau_2(\tau)-\tau$ (solid line) and $\tau_3(\tau)-\tau$ (dashed
line) on $\tau$ for disturbed rotation}
 \label{Ygr}\end{figure}

In Ref.~\cite{stab} we investigated numerically disturbed
rotational states of the string configuration Y and observed
instability of the states (\ref{roty}). Omitting details of
numerical modelling (described in Ref.~\cite{stab}), we
demonstrate in Fig.~\ref{Ygr} some ``photographs'' of the rotating
Y configuration with constant time intervals and dependence of
deviations $\tau_2(\tau)-\tau$ (solid line) and
$\tau_3(\tau)-\tau$ (dashed line) on the time parameter $\tau$ for
disturbed rotational states (\ref{roty}).
 Here we test the state with masses (\ref{m123}) for $\theta=0.1$ and
 with small initial disturbance of the component $\dot\Psi_{1\pm}^1(\tau)$.
During further evolution small disturbances grow, the junction
moves, lengthes of three arms vary and at last one of massive
points merge with the junction. Numerical experiments demonstrate
that evolution of small disturbances for velocities $U_j^\mu$ or
values $\tau_j(\tau)$ corresponds to expression (\ref{linmod}),
amplitudes of disturbances linearly grow and frequency of
oscillations (with respect to $\tau$) is equal $\theta$.

This behavior lets us to conclude, that rotational states
(\ref{roty}) of the string model Y  are unstable, because an
arbitrary small disturbance contains linearly growing modes of the
type (\ref{linmod}) in its spectrum.

If we compare these two types of instability: exponential growth
(\ref{expo}) of small disturbances for the linear string model
$q$-$q$-$q$ and linear  growth (\ref{linmod}) for the Y string
model, we are to make the following conclusion. Instability of the
model $q$-$q$-$q$ has the characteristic time $t_{inst}$ and
correspondent width $\Gamma_{inst}\simeq1/t_{inst}$. But for the Y
model with linear  growth (\ref{linmod}) we have no any
characteristic time, this corresponds to zero contribution
$\Gamma_{inst}=0$ in the increment of instability. So instability
of rotational states (\ref{roty}) does not give an additional
contribution in width $\Gamma$ of baryons, described with the Y
string model.

This rotational instability is not a weighty argument against
application of the Y configuration. But this model has another
drawback mentioned above, it predicts the slope
$\alpha'=(3\pi\gamma)^{-1}$ for Regge trajectories, that differs
from the value $\alpha'=(2\pi\gamma)^{-1}$ for the string with
massive ends \cite{Ko,4B}. The experimental value of Regge slope
$\alpha'\simeq0.9$ GeV${}^{-2}$ is close for mesons and baryons.
So the effective value of string tension $\gamma$ is to be
different for the baryon models Y  and the meson model
$q$-$\overline{q}$.

These arguments work in favor of the quark-diquark model
(Fig.~\ref{mod}{\it b}) for describing baryons on Regge
trajectories. In the next section we consider the common scheme
for describing these trajectories  for mesons and baryons.

\section{Strings and Regge
trajectories for mesons and baryons}\label{Regge}

Rotational states (\ref{lrot}) of the string with massive ends
were applied for describing excited mesons and baryons on Regge
trajectories in Refs.~\cite{4B,Ko,PRTr,InSh,Solovm}. All mentioned
authors used quasilinear dependence between angular momentum $J$
and square of energy $E$  of a  state (\ref{lrot}).

Expressions for energy $E$ (or mass $M=E$) and angular momentum
$J$ of a rotational state (\ref{lrot}) for the string with massive
ends have the following form \cite{4B,InSh}:
 \begin{eqnarray}
M=E&=&\frac{\pi\gamma\theta}\Omega+\sum_{j=1}^2
\frac{m_j}{\sqrt{1-v_j^2}}+\Delta
E_{SL}, \label{E}\\
 J=L+S&=&
\frac{1}{2\Omega}\bigg(\frac{\pi\gamma\theta}\Omega+\sum_{j=1}^2
\frac{m_jv_j^2}{\sqrt{1-v_j^2}} \bigg)+\sum_{j=1}^ns_j.\qquad
 \label{J}\ \end{eqnarray}
 Here $s_j$ are spin projections of massive points, $\Delta E_{SL}$
is the spin-orbit contribution to the energy in the following form
\cite{4B,InSh}:
 \begin{equation}
\Delta E_{SL}=\sum_{j=1}^2 \beta(v_j)( \textbf{$\Omega$}\cdot
\textbf{s}_j),\quad \beta(v_j)=1-(1-v_j^2)^{1/2}.
 \label{corr} \end{equation}
 This form of the spin-orbit contribution results from the assumption
about pure chromoelectric field in the rotational center rest
frame \cite{4B,Ko,AllenOVW}. The authors of Ref.~\cite{Ko} used
the  alternative expression
 \begin{equation}
\beta(v_j)=1-(1-v_i^2)^{-1/2},
 \label{corrT}\end{equation}
 corresponding to the Thomas
precession of the spins $\textbf{s}_j$.

If the string tension $\gamma$, values $m_j$ and $s_j$  are fixed,
we obtain an one-parameter set of rotational states (\ref{lrot}).
Values $J$ and $E^2$ for these states form the quasilinear Regge
trajectory with asymptotic behavior
 $ J\simeq\al'E^2$  for large $E$ and $J$ \cite{4B} with the slope
 $\al'=1/(2\pi\gamma)$.

We use the model of string with massive ends (considered as the
model $q$-$\overline{q}$ of a meson and the quark-diquark model
$q$-$qq$ of a baryon) to describe experimental data for excited
states of mesons and baryons on Regge trajectories. For this
purpose we are to choose free parameters of the model: effective
value of string tension $\gamma$ and effective masses of quarks
and diquarks  $m_j$ for all flavors.

This approach was developed in Refs.~\cite{4B,InSh}, but in the
present paper we study both types of spin-orbit correction (\ref{corr})
and (\ref{corrT}), use the optimization procedure for choosing the
mentioned effective values $\gamma$, $m_j$  and also include
charmed and bottom hadrons. The main principle of this choice is
to describe the whole totality of experimental data on excited
mesons and baryons \cite{PDG12}.

%
%

\begin{figure}[bh]
\includegraphics[scale=0.63,trim=17 5 10 20]{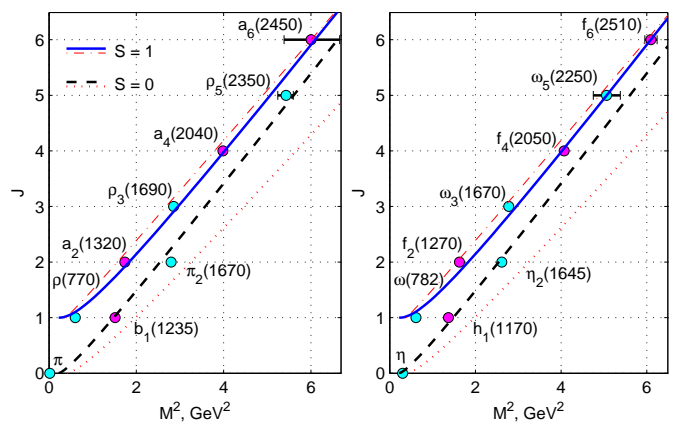}
\caption{Regge trajectories with corrections (\ref{corr}) (heavy lines)
and (\ref{corrT}) (thin lines) (a) for
isovector mesons $\rho$, $a$, $\pi$; (b)  for  isoscalar mesons
$\omega$, $f$, $\eta$.}
 \label{mes1}\end{figure}

At the first stage we describe main  Regge trajectories for light
unflavored mesons and choose effective values of tension $\gamma$
and mass $m_{ud}$ of the lightest quarks $u$ and $d$ (we suppose
below that they are equal: $m_u=m_d=m_{ud}$). The results for such
isovector and isoscalar mesons with  spin-orbit corrections
(\ref{corr}) and (\ref{corrT}) are shown in Fig.~\ref{mes1}.

 Here parameters for models with Eqs.~(\ref{corr}), (\ref{corrT}) are
 \begin{table}[ht]
\caption{Effective values of parameters $\gamma$, $m_{ud}$, $m_s$ for spin-orbit corrections
(\ref{corr}) and (\ref{corrT}).}
\begin{tabular}{||c||c|c|c||}  \hline
Correction & \ $\gamma$ (GeV${}^2$) \ & \ $m_{ud}$ {MeV} \ & \ $m_s$ {MeV} \ \\ \hline
(\ref{corr}) & 0{.}154 & 231{.}5  & 369{.}0  \\ \hline
(\ref{corrT})& 0{.}1767 & 320{.}0 & 436{.}0\\
\hline
 \end{tabular}
 \label{Tab1}\end{table}


 These values are obtained in the following optimization
procedure.

We fix a set of $n$ mesons with masses $M_k$ and angular momenta
$J_k$, $k=1,\dots,n$, with the definite quark composition. These
mesons may lie on one or on a few Regge trajectories differing in
quark spin $S$ or isospin $I$. Such a set is shown  in
Fig.~\ref{mes1}, it includes 4 Regge trajectories.

For the best fitting between the model dependence $J=J(M)$
(\ref{E}), (\ref{J}) and the experimental values $M_k$, $J_k$ from
the table \cite{PDG12}  for this set of mesons we use the least-squares method and
minimize the sum of squared deviations with positive weights
$\rho_k$:
   \begin{equation}
F(m_1,m_2,\gamma)=\sum_{k=1}^n\rho_k\Big[J_k-J(M_k)\Big]^2.
 \label{Sumkv}\end{equation}

The weights $\rho_k$ correspond to data or model errors, they are fixed
below in the following manner: $\rho_k=1$ for reliable meson
states from summary tables \cite{PDG12} with orbital momenta $L\ge1$;
$\rho_k=0{.}2$ for unreliable states with high $J$ omitted from summary tables \cite{PDG12},
in particular, for $\rho_5(2350)$ in Fig.~\ref{mes1}, but $\rho_k=0{.}1$ for states
of this type, that need confirmation ($\omega_5$, $a_6$, $f_6$);
$\rho_k=0{.}2$ for states with  $L=0$. States with  $L=0$, in particular,  $\pi$,
$\rho(770)$, $\eta$, $\omega(782)$ in Fig.~\ref{mes1} should not be described by string models,
because the string shape may correspond only to extended hadron states with high $J$ \cite{Nambu,Ch,AY,4B,Ko}.

\begin{figure}[hb]
\includegraphics[scale=0.58,trim=14 8 10 10]{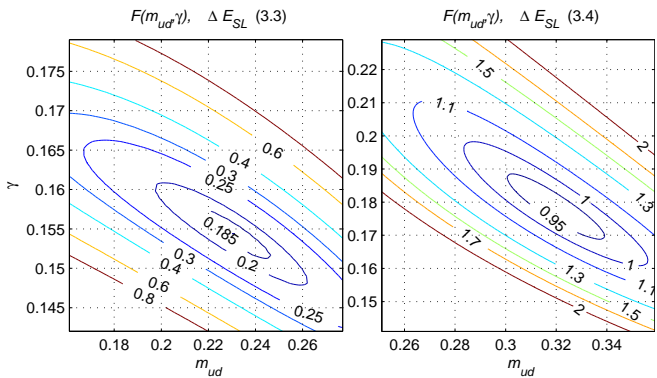}
\caption{Level lines of the
sum (\ref{Sumkv}) $F(m_{ud},\gamma)$ for spin-orbit correction (\ref{corr})
 (left) and (\ref{corrT}) (right).}
 \label{meslin}\end{figure}

To determine theoretical values of angular momenta $J(M_k)$, corresponding to
masses $M_k$ from the table \cite{PDG12}, we are to invert numerically the function $M(\Omega)$
(\ref{E}), (\ref{v1}) and substitute the function $\Omega=\Omega(M)$ into Eq.~(\ref{J}):
$J(M_k)=J\big(\Omega(M_k)\big)$.
We calculate the sum (\ref{Sumkv}) for the mesons in Fig.~\ref{mes1}
with both types of spin-orbit correction (\ref{corr})
and (\ref{corrT}) for different effective values $\gamma$ and $m_{ud}$ (here $m_1=m_2=m_{ud}$).
The results of this calculation are presented in  Fig.~\ref{meslin} as level lines of the
function (\ref{Sumkv}) in the $(m_{ud},\gamma)$ plane.

One can see that the sum (\ref{Sumkv}) for the model (\ref{corr})
reaches its minimum $F_{min}\simeq0{.}18$, if the effective parameters $m_{ud}$, $\gamma$
are close to the values in Table~\ref{Tab1}. The similar minimum $F_{min}\simeq0{.}94$ for the model
with correction (\ref{corrT}) is 5 times larger. So this model is less successful
in describing these meson states (thin lines in  Fig.~\ref{mes1}).

The results in Table~\ref{Tab1} are obtained with taking into account
8 leading Regge trajectories of mesons: to 4 mentioned above trajectories
we add 4 sets of mesons with the strange quark $s$. They include  2 sets of
$K$ mesons with $m_1=m_{ud}$ and $m_2=m_s$, in particular, $K$, $K_1(1270)$, $K_2(1770)$
with summary quark spin $S=0$ and $K^*(892)$, $K^*_2(1430)$,
$K^*_3(1780)$, $K^*_4(2045)$, $K^*_5(2382)$ with $S=1$.
The sets $\eta$, $h_1(1380)$, $\eta_2(1870)$ with $S=0$ and
 $\phi(1020)$, $f'_2(1525)$, $\phi_3 (1850)$ with $S=1$ are supposed to be
 $s$-$\overline s$ mesons: $m_1=m_2=m_s$. The weights $\rho_k$
in Eq.~(\ref{Sumkv}) are determined as mentioned above, for example,
$\rho_k=0{.}2$ for $K$, $K^*(892),\dots h_1(1380)$, $\rho_k=0{.}1$ for
$K^*_5$ and $\eta_2$.

The sum (\ref{Sumkv}) for these 8 Regge trajectories with 32 mesons depends on 3 parameters:
$F=F(m_{ud},m_s,\gamma)$. The minimal values $F_{min}\simeq0{.}301$  for the model (\ref{corr})
and $F_{min}\simeq1{.}267$ (4 times larger) for the case (\ref{corrT}) are reached, if these 3 parameters
take the values in Table~\ref{Tab1}.
for the model (\ref{corr}) these values  differ from the effective parameters
used in Ref.~\cite{4B} $\gamma=0{.}175$ GeV${}^2$, $m_{ud}=130$ MeV, $m_s=270$ MeV.
The values from Table~\ref{Tab1} essentially diminish the
sum $F(m_{ud},m_s,\gamma)$.

\begin{figure}[th]
\includegraphics[scale=0.63,trim=23 5 10 20]{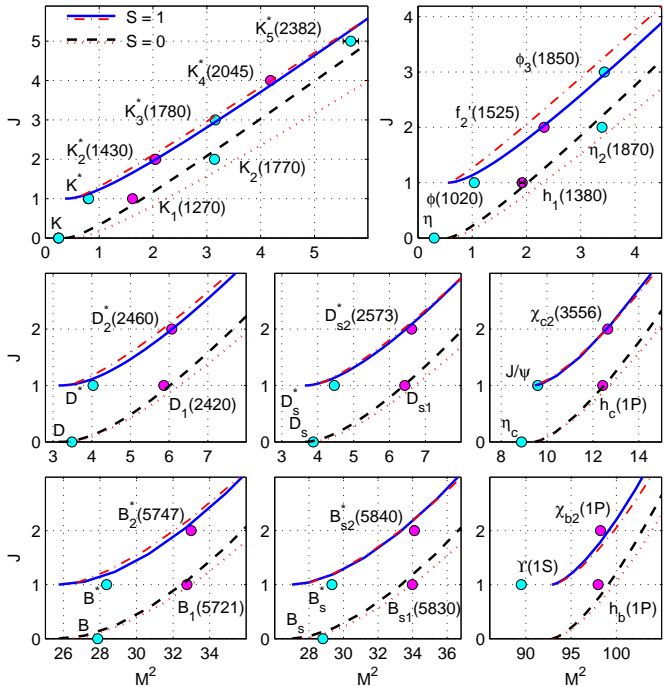}
\caption{Regge trajectories for strange, charmed, bottom mesons
with model parameters $\gamma$, $m_q$ from Tables~\ref{Tab1},
\ref{Tab2}.}
 \label{mes2}\end{figure}

Description of 4 Regge trajectories of mesons with $s$ quark is
presented in Fig.~\ref{mes2}. Here we use the same notations:
heavy solid lines for the case (\ref{corr}) $S=1$, dashed lines
for $S=1$, thin dash-dotted and dotted lines for the case
(\ref{corrT}). One can see, that the model with spin-orbit
contribution (\ref{corr}) demonstrate better agreement.

If parameters $\gamma$, $m_{ud}$, $m_s$  of the model are fixed in
Table~\ref{Tab1}, we can determine the best value $m_c$ of $c$
quark for describing $D$ mesons and charmonium states. They form 6
Regge trajectories: $D$, $D_1(2420)$, $D^*$, $D^*_2(2460)$ are
charmed analogs of $K$ and $K^*$ mesons; $D_s$, $D_{s1}(2536)$,
$D_s^*$, $D^*_{s1}(2573)$ are their partners $c\overline s$ or
$s\overline c$; $\eta_c$, $h_c(1P)$, $J/\psi$, $\chi_{c2}$ are
$c\overline c$ states. As stated above we take $\rho_k=0{.}2$ for
 mesons with $L=0$ and $\rho_k=1$ for $L=1$.
Under these  circumstances we minimize the function  (\ref{Sumkv})
$F=F(m_c)$ of one argument and obtain the optimal values $m_c$ in
Table~\ref{Tab2} for both considered models with spin-orbit
corrections (\ref{corr}) and (\ref{corrT}).

 \begin{table}[bh]
\caption{Effective values of parameters $m_c$, $m_b$.}
\begin{tabular}{||c||c|c||}  \hline
Correction  &  \ $m_c$ {GeV} \ & \ $m_b$ {GeV} \ \\ \hline
(\ref{corr})  & 1{.}5372  & 4{.}818  \\ \hline
(\ref{corrT}) & 1{.}5322 & 4{.}8198\\
\hline
 \end{tabular}
 \label{Tab2}\end{table}

Six Regge trajectories for charmed mesons are presented in
Fig.~\ref{mes2}. Note that for them we used only one fitting
parameter $m_c$, but all trajectories are described rather well in
both models with Eqs.~(\ref{corr}) and (\ref{corrT}).

The similar approach to bottom mesons $B$, $B^*$, $B_s$, $B^*_s$,
$\Upsilon$, $\chi_{b2}(1P)$ with substitution $c$  for $b$ quark
results in the optimal values $m_b$ for bottom quark in
Table~\ref{Tab2} and corresponding Regge trajectories for bottom
mesons in Fig.~\ref{mes2}. Here we use the notations of
Fig.~\ref{mes1}, in particular, thick lines correspond to
spin-orbit correction (\ref{corr}). This model has advantage in
comparison with the case (\ref{corrT}) (thin lines) for mesons
with strange quark $s$. For charmed and bottom mesons this
advantage becomes inessential. Both models are successful in
describing charmonium states, but they work worse for bottomonium.

The similar approach may be applied to baryons. We mentioned above
that for describing baryonic Regge trajectories we are to choose
the quark-diquark string model. Only this model predicts
rotational stability  and natural Regge slope
$\alpha'=(2\pi\gamma)^{-1}$ corresponding to equal experimental
Regge slopes $\alpha'$
 both for mesons and baryons.

When can apply to baryons the model of a string with massive ends,
if we determine the effective diquark mass $m_d=m_2$ (assume that
$m_1$ is the quark mass). In the simplest approach \cite{4B} we
suppose that diquarks are weakly bound systems, so a diquark mass
is close to sum of two constituent quark masses, in particular,
for $N$ and $\Delta$ baryons $m_d\simeq2m_{ud}$.

Another approach is widely used in string and potential
quark-diquark models of baryons \cite{Ko,KlemptR09,EbertFauG11}
and supposes different diquark masses for scalar diquarks with
total spin $S_d=0$ and for vector diquarks with $S_d=1$. Essential
difference of these masses corresponds to strong coupling between
two quarks in a diquark. However this mechanism remains vague, the
diquark masses in the mentioned models are used as fitting
parameters. In particular, masses $m_d^0$ of the scalar $[u,d]$
diquark in $N$ baryons and $m_d^1$ for the vector $\{u,d\}$
diquark in $\Delta$ baryons in the potential model
\cite{EbertFauG11} are correspondingly 710 and 909 MeV. The
similar masses in the string model~\cite{Ko} are $m_d^0=220$ and
$m_d^1=550$ MeV. This difference requires special explanation.

The last approach may be applied to our string model with two
forms (\ref{corr}) and (\ref{corrT}) of spin-orbit correction. For
this purpose we use two sets of baryonic Regge trajectories with
scalar and vector diquarks correspondingly. For scalar diquarks we
choose the main Regge trajectory for $N$ baryons $N$, $N(1520)$,
$N(1680)\dots$ and the corresponding trajectory for $\Lambda$
baryons: $\Lambda$, $\Lambda(1520)$, $\Lambda(1820)\dots$ If we
use data for mesons from Table~\ref{Tab1}, fix tension $\gamma$,
the mass $m_1=m_{ud}$ or $m_s$ of a single quark and vary the free
parameter $m_d=m_d^0$, we determine the optimal value $m_d^0$ for
describing the mentioned Regge trajectories with the minimal sum
(\ref{Sumkv}). The similar optimal values $m_d^1$ for the vector
$\{u,d\}$ diquark with $S_d=1$ is calculated on the base of
trajectories with $\Delta$ baryons $\Delta(1232)$, $\Delta(1930)$,
$\Delta(1930)\dots$ and $\Sigma$ baryons $\Sigma(1385)$,
$\Sigma(1775)$, $\Sigma(2030)$. Optimal values of these scalar and
vector diquark masses {MeV} are presented in Table~\ref{Tab3}.

 \begin{table}[ht]
\caption{Diquark masses  $m_d$ for models (\ref{corr}) and
(\ref{corrT}).}
\begin{tabular}{||c||c|c||}  \hline
Correction &  $m_d^0$  for $S_d=0$ &  $m_d^1$ for $S_d=1$  \\
\hline (\ref{corr}) & 412{.}6  & 588  \\ \hline
(\ref{corrT}) & 352{.}9 & 702\\
\hline
 \end{tabular}
 \label{Tab3}\end{table}

One can see that in the model with spin-orbit correction
(\ref{corrT}) optimal masses of scalar and vector diquarks are
essentially different: $m_d^1\simeq2m_d^0$. This difference
corresponds to so the similar relation between  $m_d^1$ and
$m_d^0$ in Ref.~\cite{Ko}, where the same correction (\ref{corrT})
in the quark-diquark model was used.

This difference is not so large in the model with spin-orbit
correction (\ref{corr}), the value $m_d^0$ in Table~\ref{Tab3} is
close to $2m_{ud}=463$ MeV. So in this model we can use not only
optimal parameters from Table~\ref{Tab3}, but also $m_d=2m_{ud}$.

\begin{figure}[th]
\includegraphics[scale=0.65,trim=12 5 10 20]{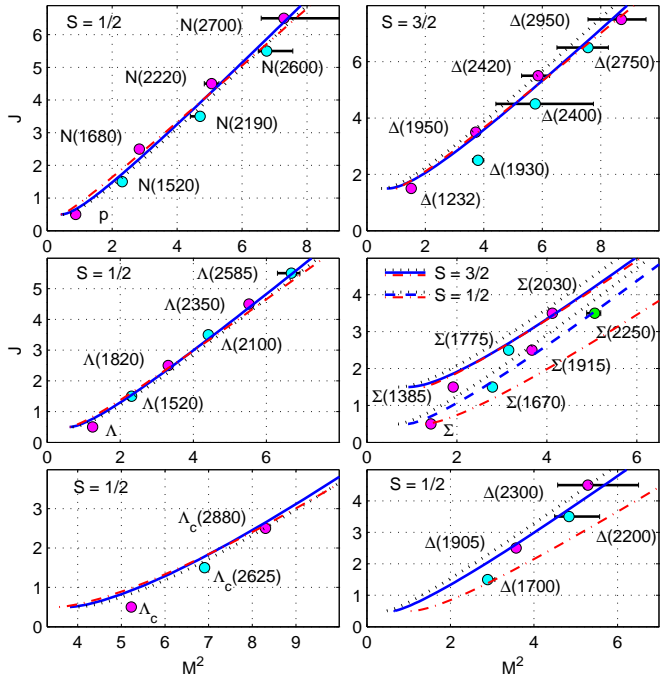}
\caption{Regge trajectories for baryons in models (\ref{corr})
 (solid  and dotted lines) and (\ref{corrT}) (thin dash-dotted lines).}
 \label{bar1}\end{figure}

Regge trajectories for the mentioned baryons are shown in
 Fig.~\ref{bar1}. The model parameters for both models (\ref{corr}) and
(\ref{corrT}) are from Tables~\ref{Tab1}\,--\,\ref{Tab3}.
Notations are similar to  Fig.~\ref{mes2}, but for the model
(\ref{corr}) we draw predictions with $m_d^0=412{.}6$ and
$m_d^1=588$ MeV from Table~\ref{Tab3} as thick solid lines (for
$\Sigma$ baryons the solid line describes the case with quark's
spin $S=3/2$ and the thick dashed line corresponds to $S=1/2$ with
$s_1=-1/2$, $s_2=1$). Predictions of the model (\ref{corr}) with
$m_d=2m_{ud}=463$ MeV are described with dotted lines, predictions
of the model (\ref{corrT}) with $m_d^0$ and $m_d^1$ from
Table~\ref{Tab3} are described with dash-dotted lines.

We see that for main (parent) Regge trajectories for $N$,
$\Lambda$ baryons with $S=1/2$ and for $\Delta$ and $\Sigma$
baryons with $S=3/2$ predictions of both models (\ref{corr}) and
(\ref{corrT}) with $m_d^j$ from Table~\ref{Tab3} are very close.
For $\Delta$ and $\Sigma$ baryons with $S=3/2$ these curves
practically coincide.

The similar coincidence takes place for charmed $\Lambda_c$
baryons (only these  charmed baryons form an appreciable Regge
trajectory). The trajectory for $\Lambda_c$ baryons is used here
as a test for these models, all their parameters $\gamma$, $m_c$,
$m_d^0$ were determined previously.

But for $\Sigma$ baryons $\Sigma$, $\Sigma(1670)$, $\Sigma(1915)$
and $\Delta$ baryons $\Delta(1700)$, $\Delta(1905)\dots$ the
models (\ref{corr}) and (\ref{corrT}) predict different Regge
trajectories, if we interpret these hadrons as states  with
$s_1=-1/2$, $s_2=S_d=1$. In this approach only the model with
correction (\ref{corr}) works successfully, in the case
(\ref{corrT}) the mass correction $\Delta E_{SL}$ appears to be
positive and too large. So we can describe these Regge
trajectories in the model (\ref{corrT}) only if we suppose that
diquarks in these hadrons are scalar ones. In this case
trajectories for $\Delta$ and $\Sigma$ baryons with $S=1/2$ will
be copies of trajectories for $N$ and $\Lambda$ baryons.

Dotted lines for all baryons in Fig.~\ref{bar1} show that  the
model with correction (\ref{corr}) admits the diquark mass
$m_d=2m_{ud}$. This assumption works rather good for  $N$,
$\Lambda$ and $\Lambda_c$ baryons and it works worse for $\Delta$
and $\Sigma$ baryons. Note that the model with correction
(\ref{corrT}) is incompatible with the assumption $m_d=2m_{ud}$.

\section{Conclusion}

Different string hadron models are considered from the point of
view of their application to describing Regge trajectories for
mesons and baryons. For this purpose we study the stability
problem for classical rotational states of these models. It is
shown that these states are unstable for the Y string baryon model
(Fig.~\ref{mod}{\it d}). The type of this instability differs from
that for the linear string baryon model $q$-$q$-$q$. For the Y
configuration small disturbances grow linearly, whereas for the
linear model they grow exponentially. This results in too large
additional width $\Gamma$ of excited baryons in the linear model
\cite{PR09}.

For the Y string baryon model we have no additional width, but
this model predicts the slope $\alpha'=(3\pi\gamma)^{-1}$ for
Regge trajectories, that differs from $\alpha'=(2\pi\gamma)^{-1}$
for the string with massive ends \cite{Ko,4B}. So for describing
both mesons and baryons with almost equal experimental value of
$\alpha'$ we have to use the string with massive ends as the meson
model $q$-$\overline{q}$ and as the quark-diquark baryon model
$q$-$qq$.

These models with  spin-orbit correction in two forms (\ref{corr})
and (\ref{corrT}) can describe main Regge trajectories for light
unflavored mesons, for $K$, $D$, $D_s$, $B$, $B_s$ mesons,
charmonium and bottomonium states, and also for $N$, $\Delta$,
$\Sigma$, $\Lambda$ and $\Lambda_c$ baryons. In this approach we
use the optimization procedure with choosing the effective string
tension $\gamma$ and effective masses of quarks and diquarks $m_j$
for all flavors (see Tables~\ref{Tab1}\,--\,\ref{Tab3}).

\end{document}